\documentclass[3p,times]{elsarticle}
\usepackage{ecrc}
\volume{00}
\firstpage{1}
\journalname{Nuclear Physics A}
\runauth{}
\jid{nupha}
\usepackage{amssymb}
\usepackage{lineno}
\usepackage[figuresright]{rotating}

\begin{document}

\begin{frontmatter}

\dochead{}

\title{Measurement of electrons from heavy-flavour decays in pp and Pb-Pb collisions with ALICE at the LHC}
\author{MinJung Kweon\footnote[1]{Email address: minjung@physi.uni-heidelberg.de}  for the ALICE Collaboration}
\address{Physikalisches Institut, Universit\"{a}t Heidelberg,\\
Heidelberg, 69120, Germany}

\begin{abstract}
The ALICE experiment has measured at mid-rapidity electrons from
heavy-flavour decays in pp collisions at \mbox{$\sqrt{s}$ = 2.76} and 7 TeV,
and in Pb-Pb collisions at \mbox{$\sqrt{s_{\rm NN}}$ = 2.76 TeV}.
In pp collisions, electrons from charm-hadron and from beauty-hadron decays are 
identified by applying cuts on displaced vertices.
Alternatively the relative yield of electrons from beauty decays to
those from heavy-flavour decays is extracted using electron-hadron correlations.
Results are compared to pQCD-based calculations.
In Pb-Pb collisions, the $p_{\rm T}$ dependence of the nuclear modification
factor of electrons from heavy-flavour decays is presented in two centrality classes.
The status of the analysis of  electrons from beauty decays in Pb-Pb collisions is reported, 
in view of the measurement of the corresponding nuclear modification factor.
\end{abstract}


\end{frontmatter}

\section{Introduction}
In ultra-relativistic heavy-ion collisions, heavy quarks are, due to their large mass, mainly produced 
via initial hard parton scatterings. Therefore, heavy-flavour particles are sensitive to the full evolution 
of the strongly-interacting partonic medium expected to be formed in the collisions \cite{sqgpform}.
According to perturbative QCD (pQCD), the energy-loss of a parton in the medium
depends on its colour charge and on the quark mass \cite{phymotive}. The modifications of heavy-flavour particle 
momentum distributions in Pb-Pb collisions with respect to the ones in pp collisions represent 
a sensitive probe for the medium properties and the underlying mechanism of in medium parton energy loss. 
The measurement of heavy-flavour production in pp collisions provides 
a precision test of pQCD calculations and a crucial baseline for the interpretation of the results in Pb-Pb collisions.

The nuclear modification factor, $R_{\rm AA}$, is defined as $R_{\rm AA}(p_{\rm T})$ = $\frac{1}{<T_{\rm AA}>}
\frac{{\rm d}N_{\rm AA}/{\rm d}p_{\rm T}}{{\rm d}\sigma_{\rm pp}/{\rm d}p_{\rm T}}$, where $<T_{\rm AA}>$ is the average 
nuclear overlap function for a given collision centrality class, ${{\rm d}N_{\rm AA}/{\rm d}p_{\rm T}}$ and 
${{\rm d}\sigma_{\rm pp}/{\rm d}p_{\rm T}}$ correspond to the $p_{\rm T}$-differential electron yield in nucleus-nucleus 
collisions and the $p_{\rm T}$-differential cross section in pp collisions respectively. 

The ALICE experiment \cite{jinst} is the dedicated heavy-ion experiment at the LHC. The ALICE detector has electron
identification capability down to very low $p_{\rm T}$ at mid-rapidity which allows us to study heavy-flavours via their semi-electronic decays
(branching ratio $\sim$ 10\%). The central barrel detectors of ALICE have a very good spatial resolution to separate secondary vertices.
Thus, electrons produced from beauty decays can be separated from those originated from charm decays.

\section{Analysis}
Tracks were reconstructed in the central rapidity region with the Time Projection Chamber (TPC) and the Inner Tracking System (ITS). 
Track momentum resolution is better than 4\% for \mbox{$p_{\rm T} <$ 20 GeV/$c$} and the resolution of the transverse 
impact parameter d$_{\rm 0}$\footnote[2] {The minimum distance of the track to the primary collision vertex in the plane 
perpendicular to the beam axis} is better than 75 $\mu$m for $p_{\rm T}$ $>$ 1 GeV/$c$ \cite{hfqm2011}. 
For the tracks which fulfill the track quality criteria, electron selection cuts \cite{alicehfe} were applied using the signals in the TPC, 
the Time-Of-Flight detector (TOF), the Transition Radiation Detector (TRD) and the Electromagnetic Calorimeter (EMCal). 
TPC and TOF were used for electron selection in Pb-Pb data. For pp data, two different  strategies were used. 
In addition to TPC, the first one employs the TOF and the TRD whereas the second one makes use of the EMCal.

Tracks compatible with the electron hypothesis within 3 $\sigma$ from the time of flight measured by the TOF were selected, 
thus rejecting kaons up to a momentum of $\sim$ 1.5 GeV/$c$ and protons up to $\sim$ 3 GeV/$c$. Using charge deposited in the TRD, 
the electron likelihood of a track was calculated. A momentum dependent cut was applied on the likelihood to have a constant 
electron efficiency of 80\%, providing excellent separation of electrons from pions up to \mbox{8 GeV/$c$} for pp collisions.
Alternatively, the ratio $E/p$ of the energy deposited in the EMCal and the measured momentum was calculated, and tracks within 
3 $\sigma$ from the $E/p$ peak were selected.  Further hadron rejection was done by applying a cut on TPC specific energy deposit 
expressed as the distance to the expected energy deposit of electrons, normalized by the energy-loss resolution.
The remaining hadron contamination, which is below 3 (10)\% up to \mbox{8 (6) GeV/$c$} for pp (Pb-Pb) data, was calculated using 
a fit of the TPC d$E$/d$x$ in momentum slices and subtracted. 

The inclusive electron spectrum consists of background electrons from various sources in addition to the signal (electrons from 
heavy-flavour decays). The main background electrons are originating from Dalitz decays of light-neutral mesons and dielectron
decays of light-vector mesons, and from the conversions of decay photons in the beam pipe and the innermost layer of the ITS. 
They were calculated based on ALICE measured $\pi^{0}$ and $\eta$ spectrum \cite{alice_pieta} and $m_{\rm T}$-scaled spectra 
for other light-flavour mesons using PYTHIA \cite{pythia} decay kinematics. The background electrons from dielectron decays of 
heavy quarkonia were calculated based on measurements at the LHC \cite{alicejpsi, cmsjpsi, cmsupsilon}, and those from  real 
and virtual QCD photon decays were estimated based on Next-to-Leading Order (NLO) pQCD calculation \cite{nlophoton}. At high electron 
$p_{\rm T}$, contributions from hard scattering processes are important and become dominant \cite{alicehfe}. The $p_{\rm T}$ spectrum 
of electrons from heavy-flavour decays was obtained by subtracting the background cocktail from the inclusive electron 
$p_{\rm T}$ spectrum.   

In pp collisions, the measurement of the $p_{\rm T}$-differential cross section of electrons from the beauty decays was done 
by applying an additional cut on the d$_{\rm 0}$ of identified electron tracks \cite{aliceb2e}. Since the electrons from beauty 
decays have larger d$_{\rm 0}$ compared to those of background electrons due to their large mean proper decay length \mbox{($c\tau$ $\cong$ 500 $\mu$m),} 
a cut on the minimum d$_{\rm 0}$ of the electron candidate tracks enhances the signal to background ratio. The remaining background 
electrons were estimated based on other ALICE measurements ($\pi^{0}$ $p_{\rm T}$ spectra \cite{alice_pieta} 
and D-mesons $p_{\rm T}$ spectra \cite{alice_d2h}), and subtracted. After corrections for geometrical acceptance and efficiency, 
the electron yield from beauty decays  per minimum bias collision was normalized using the cross section of minimum bias pp collisions.

Alternatively, the relative beauty contribution to the heavy-flavour electron yields was measured based on characteristic azimuthal angular 
correlations of electrons from heavy-flavour decays and charged hadrons (e-h). It was extracted by fitting the correlation distribution 
with Monte Carlo templates obtained using PYTHIA \cite{pythia} .

\begin{figure}[h]
\begin{minipage}{18pc}
\includegraphics[width=18.1pc]{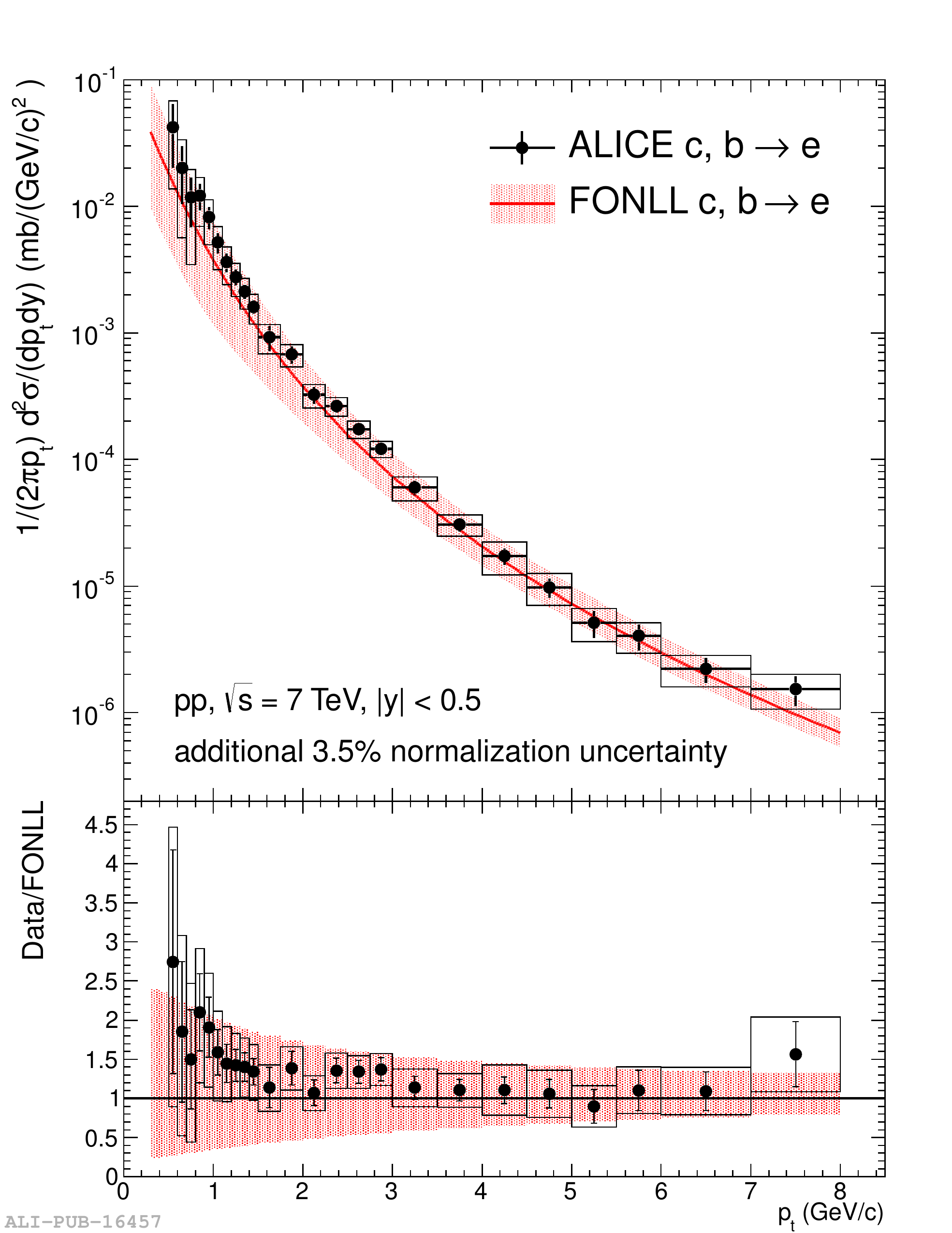}
\caption{\label{fig1}Production cross section of electrons  from heavy-flavour decays in pp collisions at  $\sqrt s$ = 7 TeV \cite{alicehfe} compared to FONLL pQCD calculations \cite{fonll}.}
\end{minipage}\hspace{2pc}
\begin{minipage}{18pc}
\vspace{0.8pc}
\includegraphics[width=16.5pc]{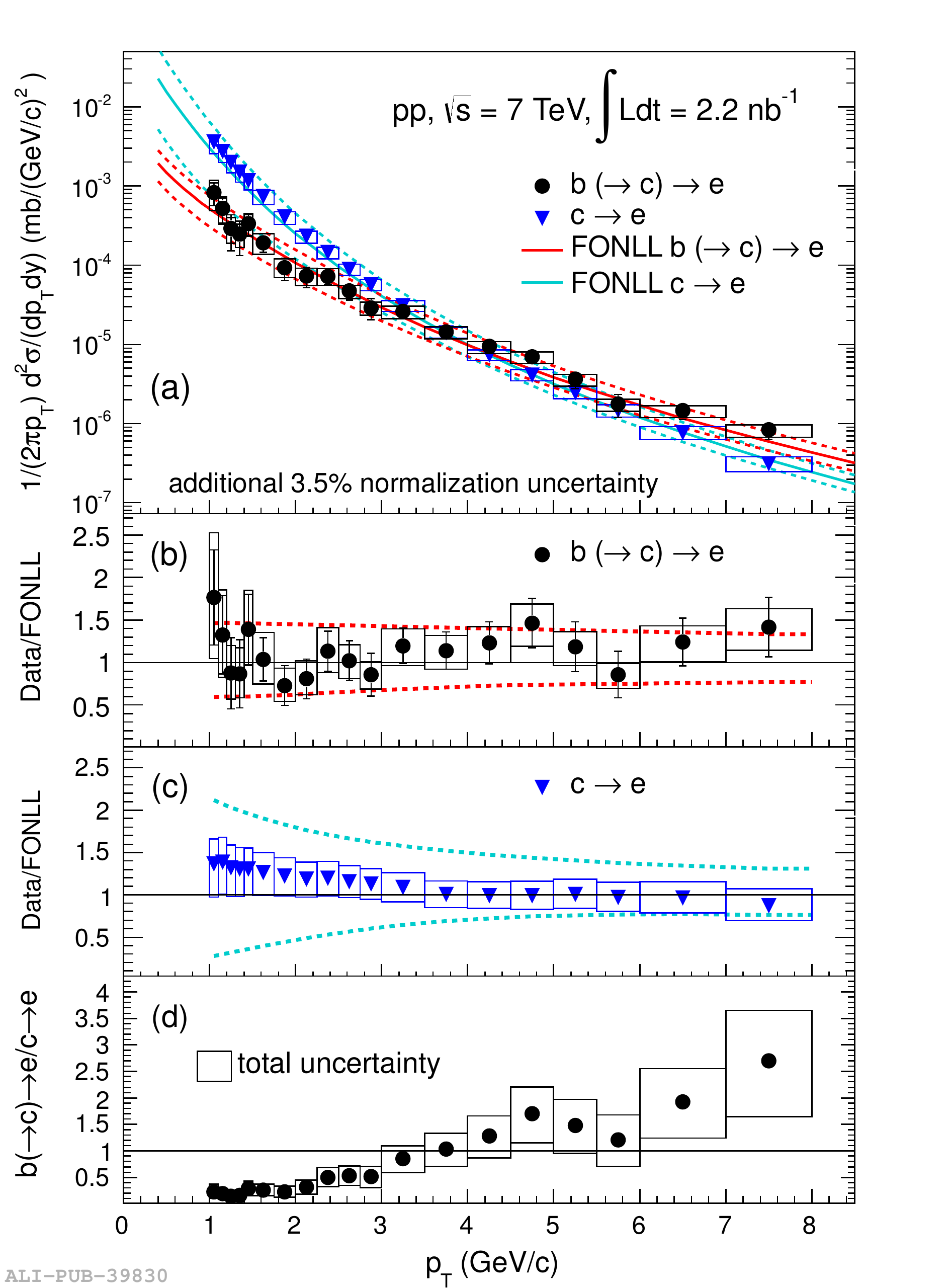}
\vspace{0.3pc}
\hspace{0.7pc}
\caption{\label{fig2} Production cross section of electrons from beauty and charm-hadron decays and their ratio  
in pp collisions at  \mbox{$\sqrt s$ = 7 TeV \cite{aliceb2e},} compared to FONLL pQCD calculations \cite{fonll}.}
\end{minipage} 
\end{figure}

\begin{figure}[h]
\begin{minipage}{19pc}
\includegraphics[width=15pc]{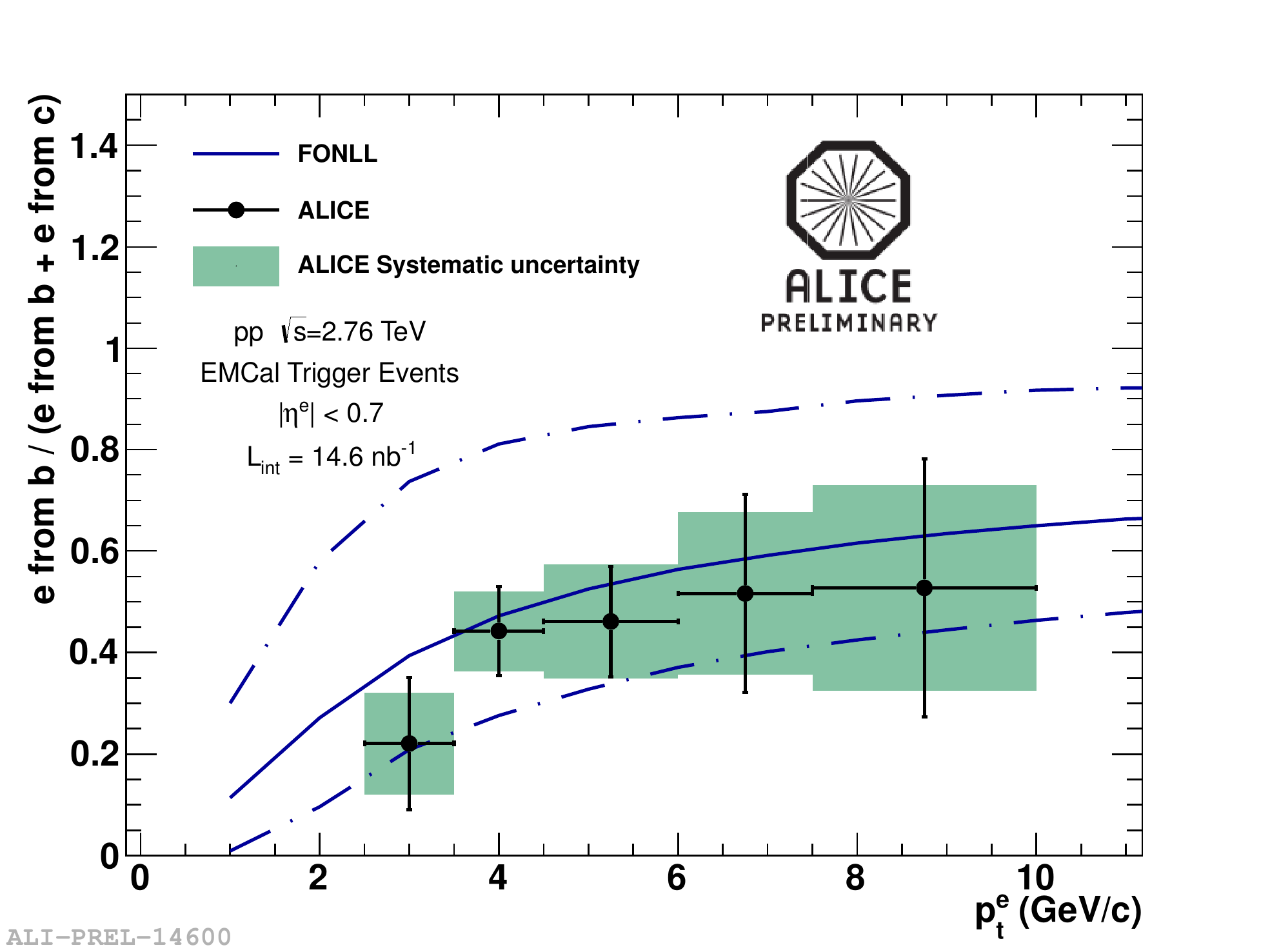}
\end{minipage}
\begin{minipage}{15.5pc}
\caption{\label{fig3}Ratio of electrons from beauty decays to those from heavy-flavour decays measured via e-h correlations, compared to FONLL calculations \cite{fonll}.}
\end{minipage} 
\end{figure}  

\section{Results in pp collisions at $\sqrt s$ = 7 TeV and $\sqrt s$ = 2.76 TeV}
The results were obtained from a sample of minimum bias pp collisions (2.6 ${\rm nb}^{-1}$ for heavy-flavour decay electrons and 2.1 ${\rm nb}^{-1}$ for beauty-decay electrons) recorded in 2010 at $\sqrt s$ = 7 TeV and a sample of EMCal triggered pp collisions ($6.2\times10^{5}$ events) recorded in 2011 at $\sqrt s$ = 2.76 TeV. 

The $p_{\rm T}$-differential invariant cross section of electrons from heavy-flavour decays is measured for $p_{\rm T}$ above \mbox{0.5 GeV/$c$} in the rapidity interval $|y|<$ 0.5 at $\sqrt s$ = 7 TeV (Figure~\ref{fig1}).
Systematic uncertainties on the measured electron spectrum amounts to $\sim$ 10($\sim$ 20)\% for $p_{\rm T}<$($>$) 3~GeV/$c$ and to $\sim$ 10($\sim$ 20)\% on the electron cocktail \cite{alicehfe}. 
Figure~\ref{fig2} presents the invariant production cross section of electrons from beauty decays in $|y|<$ 0.8 obtained with the analysis based on the d$_{\rm 0}$ cut, and the calculated electron spectrum from charm decays at $\sqrt s$ = 7 TeV. Electrons from beauty decays take over those from charm decays and become the dominant source for $p_{\rm T}$ $>$ 4 GeV/$c$.
 The relative beauty contribution to the heavy-flavour electron yields was measured using  e-h correlations at \mbox{$\sqrt s$ = 2.76 TeV} (Figure~\ref{fig3}).
In all analyses, Fixed-Order plus Next-to-Leading Log (FONLL) calculations \cite{fonll} are in agreement with the measurements within uncertainties.

\section{Results in Pb-Pb collisions at $\sqrt{s_{\rm NN}}$ = 2.76 TeV}
The results were obtained from a sample of minimum bias Pb-Pb collisions ($6.2\times10^{5}$ events) recorded in fall 2010 at $\sqrt{s_{\rm NN}}$= 2.76 TeV. 
The Silicon Pixel Detector (SPD) and the two scintillator hodoscopes (V0) provide the minimum bias trigger, and events are classified according to their centrality based on percentiles of the distribution of the sum of the amplitudes in the V0 detectors \cite{pbpbevt}. 
The inclusive electron spectra and background electron cocktails, calculated based on the charged pion spectra measured by ALICE, were obtained for six centrality classes. 
The systematic uncertainties are dominated by particle identification ($\sim$ 35\%) on \mbox{the inclusive spectra and background cocktail ($\sim$ 25\%) \cite{electronpbpb}}. 

\begin{figure}[h]
\begin{minipage}{19pc}
\includegraphics[width=15.5pc]{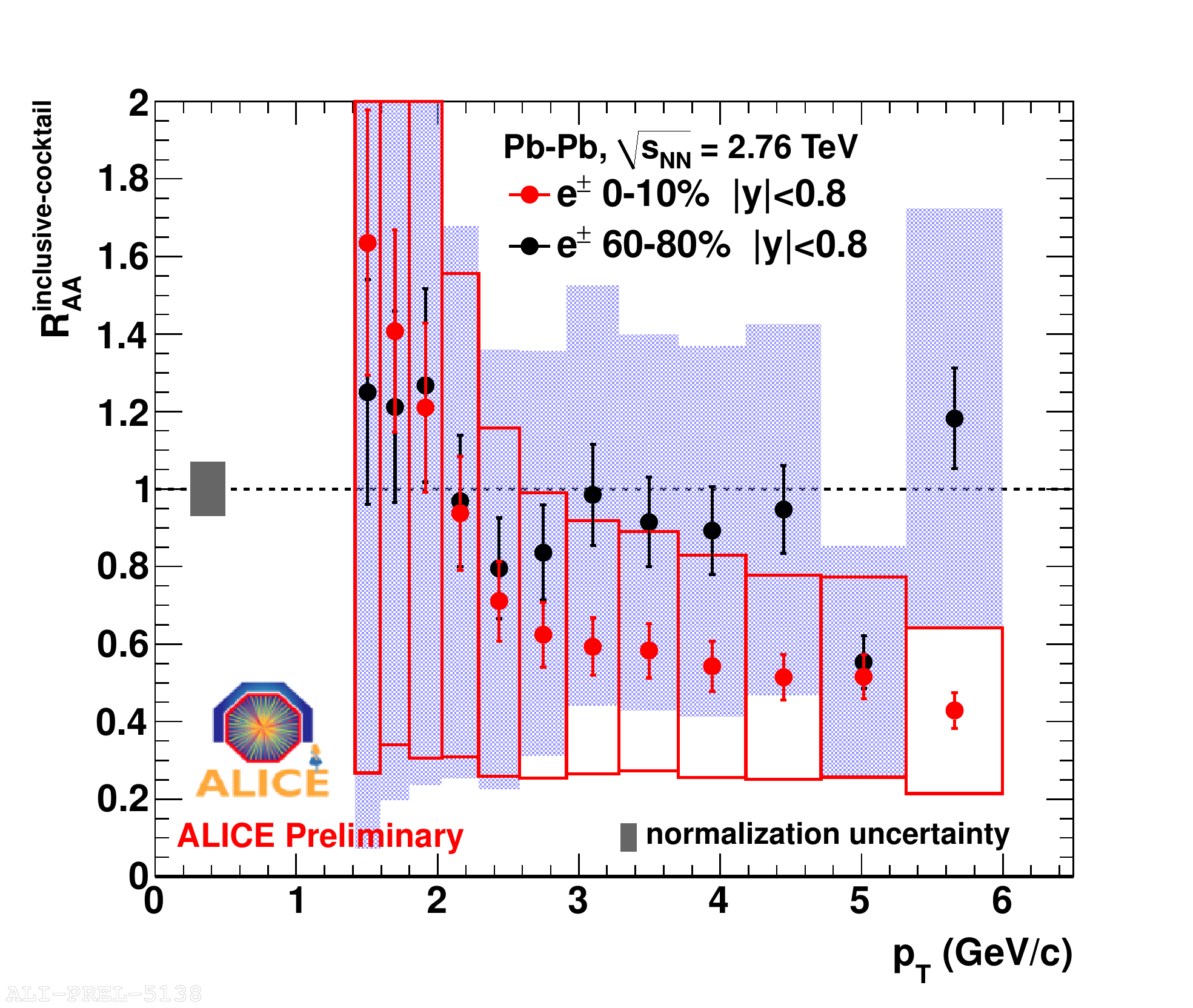}
\caption{\label{fig4}$R_{\rm AA}$ of the inclusive electron spectrum - cocktail for \mbox{0-10\%} centrality bin compared to 60-80\% centrality bin in \mbox{Pb-Pb} collisions.}
\end{minipage}\hspace{2pc}%
\begin{minipage}{20pc}
\includegraphics[width=16.7pc]{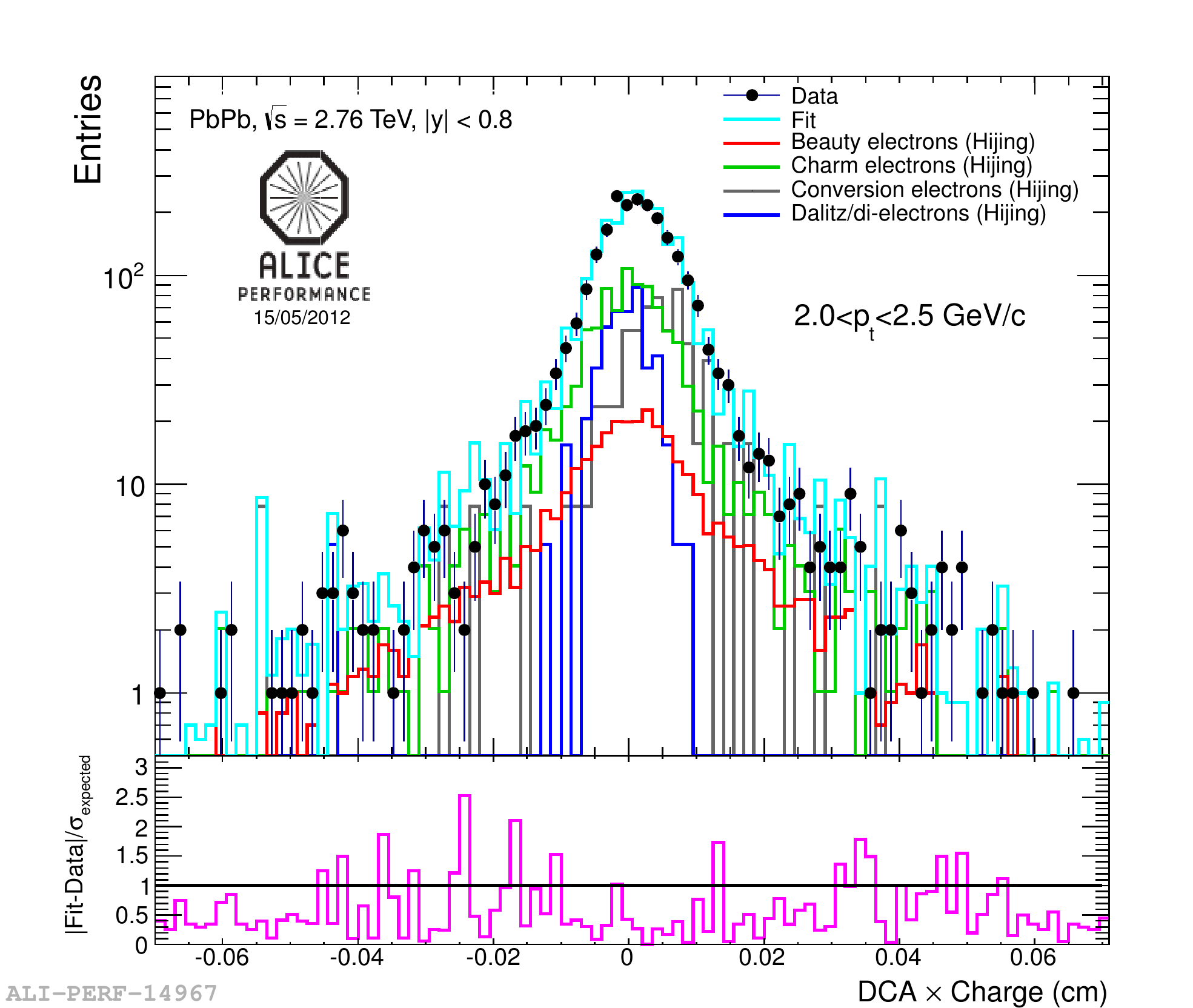}
\caption{\label{fig5}Fit of d$_{\rm 0}$ distribution using MC templates.}
\end{minipage} 
\end{figure}
The $p_{\rm T}$ dependence of the nuclear modification factor $R_{\rm AA}$ of background-subtracted electrons for  the centrality ranges 0-10\% and 60-80\% in $|y|<$ 0.8 
has been calculated with respect to pp reference spectra (Figure~\ref{fig4}).
The reference cross section in pp collisions at $\sqrt s$ = 2.76 TeV was obtained by applying a $p_{\rm T}$-dependent scaling factor, calculated based on FONLL predictions \cite{fonll}, to the cross section measured at \mbox{$\sqrt s$ = 7 TeV.}
A factor 1.5-4 suppression is observed for 3.5 $<p_{\rm T}<$ 6 GeV/$c$, where heavy-flavour decays are dominant. It suggests us a strong energy-loss of heavy quarks in the medium produced in central Pb-Pb collisions. 

The measurement of electrons from beauty decays is under study by fitting the measured d$_{\rm 0}$ distribution with Monte Carlo templates from individual sources (top of Figure~\ref{fig5}). Differences between the fit and the data are consistent with statistical variations (bottom of Figure~\ref{fig5}). Higher statistics are being analyzed for both peripheral and central collisions.
An analysis approach similar to that used for pp data is also being pursued with the Pb-Pb data. 

\section{Summary}
The $p_{\rm T}$-differential cross section of electrons from heavy-flavour decays, \mbox{and those from beauty decays}, 
were measured in pp collisions at $\sqrt{s}$ = 2.76 and 7 TeV.  
pQCD-based calculations are in agreement with measurements within uncertainties.
The electron spectra subtracted by the known background electrons were measured in \mbox{Pb-Pb} collisions at \mbox{$\sqrt{s_{\rm NN}}$ = 2.76 TeV}.
The nuclear modification factor implies strong energy-loss of heavy quarks \mbox{in the medium produced} in central Pb-Pb collisions.
The measurement of electrons from beauty decays in Pb-Pb collisions is under way. 

\bibliographystyle{elsarticle-num}

\end{document}